\documentclass[prl,twocolumn,superscriptaddress,amssymb,amsmath]{revtex4}
\usepackage{graphicx}
\usepackage{dcolumn}
\usepackage{bm}
\usepackage[latin1]{inputenc}
\usepackage[mathscr]{eucal}
\usepackage{epsfig}
\usepackage{rotating}

\newcommand{\txt}[1]{\mathrm{#1}}

\newcommand{\EC}{E_\txt{c}}
\newcommand{\kB}{k_\txt{B}}

\newcommand{\CSigma}{C_\Sigma}
\newcommand{\unit}[1]{\ \mathrm{#1}}
\newcommand{\Vol}{\mathcal{V}}
\newcommand{\VDS}{V_\mathrm{DS}}

\newcommand{\Vprobe}{V_\mathrm{probe}}

\newcommand{\Qdot}{\dot{Q}}
\newcommand{\RT}{R_\mathrm{T}}

\newcommand{\Tbath}{T_\mathrm{bath}}
\newcommand{\TS}{T_\mathrm{S}}
\newcommand{\TN}{T_\mathrm{N}}
\newcommand{\Tc}{T_\txt{c}}

\newcommand{\be}{\begin{equation}} \newcommand{\ee}{\end{equation}}
\newcommand{\ba}{\begin{eqnarray}} \newcommand{\ea}{\end{eqnarray}}

\newcommand{\ie}{i.\,e.}

\begin{document}
\preprint{ }

\title{Heat Transistor: Demonstration of Gate-Controlled Electron Refrigeration}

\author{Olli-Pentti Saira}
\author{Matthias Meschke}
\affiliation{Low Temperature Laboratory, Helsinki University of
Technology, P.O. Box 3500, 02015 TKK, Finland}

\author{Francesco Giazotto}
\affiliation{NEST CNR-INFM \& Scuola Normale Superiore, I-56126 Pisa,
Italy}

\author{Alexander M. Savin}
\affiliation{Low Temperature Laboratory, Helsinki University of
Technology, P.O. Box 3500, 02015 TKK, Finland}

\author{Mikko M\"ott\"onen}
\affiliation{Low Temperature Laboratory, Helsinki University of
Technology, P.O. Box 3500, 02015 TKK, Finland}
\affiliation{Laboratory of Physics, Helsinki University of
Technology, P.O. Box 4100, 02015 TKK, Finland}

\author{Jukka P. Pekola}
\affiliation{Low Temperature Laboratory, Helsinki University of
Technology, P.O. Box 3500, 02015 TKK, Finland}


\begin{abstract}

We present experiments on a superconductor-normal metal electron
refrigerator in a regime where single-electron charging effects are
significant. The system functions as a \emph{heat transistor}, \ie, the
heat flux out from the normal metal island can be controlled with a gate
voltage. A theoretical model developed within the framework of single-electron tunneling provides a full quantitative
agreement with the experiment. This work serves as the first
experimental observation of Coulombic control of heat transfer and, in particular,
of refrigeration in a mesoscopic system.
\end{abstract}

\maketitle

In conventional transistors such as a field-effect or
single-electron transistor~\cite{fulton87}, the electric current or
voltage is controlled by a gate. More recently, a different kind of
working principle was demonstrated in mesoscopic structures: the
electron transport can also be controlled by manipulating the
quasiparticle temperature, \ie, the energy distribution of charge
carriers~\cite{morpurgo98,baselmans99}. In these experiments, such
manipulation resulted typically in an increase of electronic
temperature. In subsequent experiments, the  supercurrent of a
superconductor-normal metal-superconductor (SNS) Josephson junction
was controlled not only by heating but also by cooling the electrons
in the active region~\cite{savin04}. A typical electron refrigerator
consists of a SNS system with two insulating barriers defining a
SINIS tunnel structure~\cite{leivo96}. Biasing the device with a
voltage of approximately twice the superconducting energy gap
results in extraction of heat from the N island, and hence a drop in
its temperature.
Ultra-small structures operated at low temperatures display
pronounced charging effects which may profoundly affect the heat
transport dynamics in mesoscopic systems.

In this Letter, we investigate a SINIS structure with very small
tunnel junctions which behaves as a gate-controlled single-electron
refrigerator. In such a system, we demonstrate experimentally gate
modulation of the \emph{heat flux} by more than a factor of three.
Our experimental findings are successfully explained through a model
which associates the thermal current with single-electron
tunneling~\cite{Pekola_Ser}. This structure provides an ideal
framework for the investigation of the interplay between charging
effects and heat transport at mesoscopic scale.

%

\begin{figure}
    \begin{center}
    \includegraphics[width=\columnwidth]{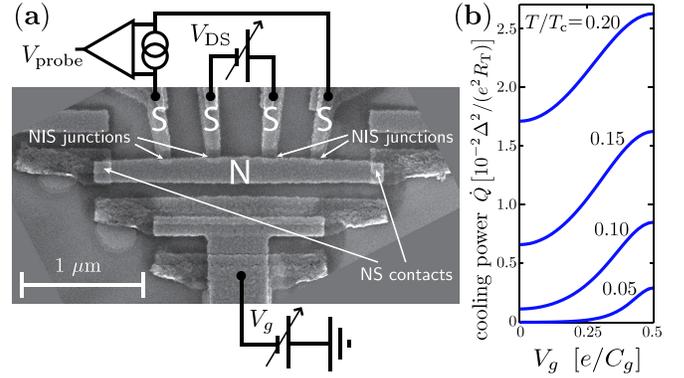}
    \end{center}
    \caption{(color online) (a) Scanning electron micrograph of the heat transistor,
    showing also the measurement setup.
    In the middle, the normal metal (Cu) island subjected to cooling.
    Four superconducting probes (Al) with Al-Ox tunnel junctions can be seen
    above, and truncated capacitor plates (Al) with direct NS
    contacts on the left and right.
    (b) Theoretical heat flow from the N island at the optimal bias point at different temperatures as a function of the gate
    voltage. $T$ denotes the common temperature of the island electrons and quasiparticles in the superconductor, and $\Tc = \Delta / (1.764 \kB)$ is the superconducting critical temperature.}
    \label{fig:sem}
\end{figure}

Figure~\ref{fig:sem}(a) shows the measured heat transistor
 with a sketch of the measurement setup. The core of the
sample consists of an N island, tunnel-coupled to four
 superconducting leads, and a gate electrode with capacitive
coupling to the island. Like in an ordinary single-electron
transistor, the charging energy penalty for tunneling electrons can
be modulated by the gate voltage. As we shall show in the following,
in the heat transistor under suitable bias conditions, Coulomb
blockade allows the modulation of both the electric current and the
heat flux. For a direct verification of the Coulomb blockade effect,
we also fabricated and measured a reference sample with the same
geometry and parameters, except for large capacitor pads (area
$5\times 10^4\,\mu$m$^2$) attached to the island with NS contacts,
thus suppressing the charging energy but ideally maintaining the
thermal isolation of the N island. In the actual heat transistor we
preserved only small superconducting protrusions instead of the
large pads in order to maintain the same topology with a minimal
effect on the charging energy. The samples were fabricated with
electron beam lithography and three-angle evaporation on an oxidized
silicon wafer. The normal metal island consists of a copper block
with lateral dimensions $180 \unit{nm} \times 2300 \unit{nm}$  and
thickness $20 \pm 3 \unit{nm}$. The superconductor metal was
aluminium, and the Al-Ox tunnel junctions had an approximate area
$120 \times 40 \unit{nm^2}$. In the heat transistor, the two inner
junctions with $110 \unit{k\Omega}$ resistances were used as a
voltage-biased cooler pair. Consistent with the transistor
terminology, we denote the bias voltage over the cooler junctions
$\VDS$. The outer junctions with resistances of approximately $150
\unit{k\Omega}$ each were current-biased at picoampere level,
enabling accurate thermometry of the island electrons by exploiting
the temperature sensitivity of a NIS junction $I$-$V$ curve in the
100-500 mK range. As only the DC properties of the device were
relevant, all signal lines to the sample stage were low-pass
filtered.

Figure~\ref{fig:sem}(b) illustrates the expected performance of the
transistor, displaying the modulation of heat flow ($\dot{Q}$) with
gate voltage ($V_g$) at different operating temperatures. The
electric and thermal transport properties of a Coulomb blockaded
SINIS structure can be derived through the theory of single-electron
tunneling assuming no exchange of energy due to environmental
fluctuations. To obtain a complete model of the experimental setup,
we consider a normal metal island with four tunnel junctions to
superconducting leads numbered \mbox{1 \ldots 4} with junction
resistances and capacitances given by $R_i$ and $C_i$, respectively,
and the electric potential of each lead denoted by $V_i$. The gate
electrode is coupled to the island with capacitance $C_g$. The total
capacitance of the normal metal island reads $\CSigma = C_g +
\sum_{i=0}^4 C_i$, where $C_0$ denotes the self-capacitance of the
island. Let us denote the total excess charge on the island by $q =
- n e$. The electrostatic energy of the system can be written in a
form where the charging energy is expressed as $E_\txt{ch} = \EC (n
+ n_g)^2$. Here $\EC \equiv e^2/(2 \CSigma)$ and $n_g = V_g C_g/e$
is the gate charge. The change in energy for an electron tunneling
to (+) or from (-) the island through junction $i$ reads
\cite{AverinLikharev}
\begin{equation}E_n^{i,\pm} = \pm 2\EC(n + n_g \pm 1/2) \pm e(V_i -
\phi), \label{eq:Epm}\end{equation} where $\phi = \frac{1}{\CSigma}
\sum_{j=1}^4 C_j V_j$ is the offset to the island potential from the
bias voltages. The $\phi$ term can be interpreted also as a shift to
the gate position $n_g$, and hence we neglect it in the following
analysis where we consider the range of gate modulation of electric
and heat current at fixed bias points. The standard expressions for
the tunneling rate $\Gamma$ and average heat flux $\dot{Q}$ as a
function of the change in energy $E_n^{i,\pm}$ for each tunneling
event are identical to those given in Ref.~\cite{Pekola_Ser}. The
charge number distribution $p_n$ is determined from the steady-state
master equation \cite{AverinLikharev}.
In the analysis of the measured data, all quantities were evaluated
numerically. The shape and position of the charge number
distribution are crucial in describing the device properties in the
Coulomb blockade regime $\kB \TN \lesssim \EC$, where $\TN$ is the N
island electron temperature.

For an analytic investigation of Coulomb blockade effects on heat
transport near the optimal bias point where the maximum cooling
power is achieved, we neglect the small probe current, and assume
the two cooling junctions to have the same resistance $R_\txt{T}$.
Furthermore, we consider the extreme cases $n_g = 0$ (closed gate)
and $n_g = -1/2$ (open gate) in the regime $\EC/(\kB \TN) \gg 1$, so
that the charge number distribution $p_n$ is symmetric and narrow.
Neglecting the population of the excited charge states, a
symmetrically biased Coulomb blockaded SINIS with an open gate
behaves as a regular SINIS whose electric and heat currents have
been scaled by $1/2$: At charge state $n=0$ ($n=1$), an electron can
enter (leave) the island through the junction biased at $\mp \VDS/2$
with no increase in charging energy, while tunneling through the
other junction is effectively blocked due to a charging energy
penalty of $2 \EC$. We can therefore apply the known results valid
in the absence of charging effects \cite{Anghel}, namely that that
the optimal cooler bias voltage is $V_\txt{DS}^\txt{opt} \simeq
2(\Delta - 0.66 \kB \TN)/e$, and the maximum cooling power for the
heat transistor reads
\begin{equation}\Qdot^\txt{opt}_\txt{open} \simeq
0.59 \frac{\Delta^{1/2} (\kB \TN)^{3/2}}{e^2 \RT}.
\label{eq:Qopen}
\end{equation}
With a closed gate, all tunneling events are affected by Coulomb
blockade and the cooling power is minimized. The leading term of the
cooling power corresponds to tunneling events occurring at the $n =
0$ state, and employing similar approximations as in
Eq.~(\ref{eq:Qopen}), the cooling power accounting for the charging
energy penalty of magnitude $\EC$ can be written
\begin{equation}
\Qdot^\txt{opt}_\txt{closed} \simeq 1.3 \frac{\EC}{\kB \TN}
\exp\left(-\frac{\EC}{\kB \TN}\right) \frac{\Delta^{1/2} (\kB
\TN)^{3/2}}{e^2 \RT}. \label{eq:Qclosed}
\end{equation}
Comparison with numerical computations shows that the above
approximative formulas are accurate to within 10\% when $\EC/(\kB
\TN) \gtrsim 2.5$.

\begin{figure}
    \begin{center}
    \includegraphics[width=\columnwidth]{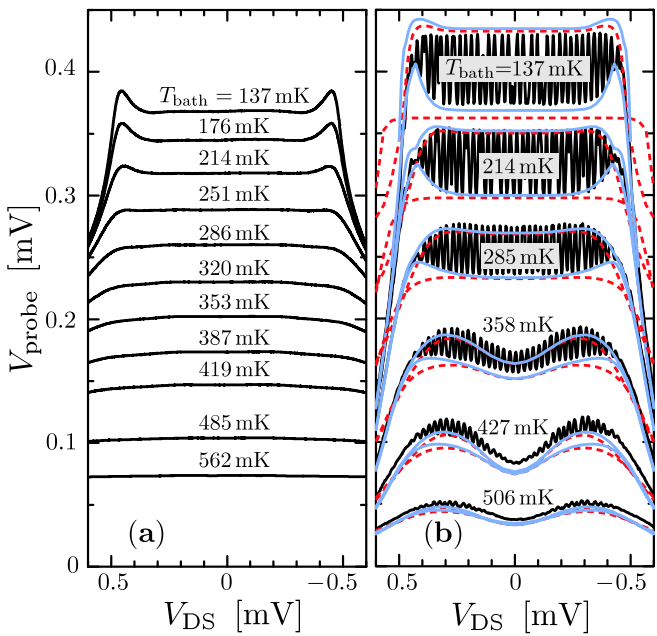}
    \end{center}
    \caption{(color online) Measured probe voltage as a function
    of the drain-source bias voltage for (a) the reference
    sample with negligibly small charging energy and (b) the heat
    transistor at different bath temperatures. For the heat transistor,
    the gate voltage was swept during the measurement, and hence the plot
    displays the full range of gate modulation at each bias point,
    whereas there was no observable gate modulation in the reference
    sample. Superimposed on panel (b) are theoretical curves for the two extremal
    gate positions. The solid blue curves are calculated assuming an electron-phonon heat
    load given by Eq.~(\ref{eq:Qeph}) with the experimentally determined value
    $\Sigma = 2.3 \times 10^9 \unit{W K^{-5} m^{-3}}$, and the dashed red curves given for
    reference show the expected behavior where the electron
    temperature  is fixed at the bath tempeature ($\Tbath$) over the full bias range.
    }
    \label{fig:data}
\end{figure}

In the experiment, we recorded the voltage over the current-biased
probe junctions as the gate voltage was swept at a rate of $\sim 1$
Hz, while the voltage over the cooler junctions was swept
significantly slower over a range of more than $2 \Delta/e$. The
sample parameters $\Delta = 230 \unit{\mu V}$ and $\CSigma =
1.5\unit{fF}$ were determined from a fit to the measured $I$-$V$
characteristics and the gate modulation amplitude of $\Vprobe$,
respectively. We charted a number of bath temperatures ranging from
50~mK to 560~mK. In Fig.~\ref{fig:data}, sweeps over cooler bias
($V_{\text{DS}}$) are shown at selected bath temperatures for both
the reference sample with vanishing charging energy
[Fig.~\ref{fig:data}(a)] and for the heat transistor
[Fig.~\ref{fig:data}(b)]. Due to very slow sweeping rate compared to
the electron-phonon relaxation time~\cite{Taskinen}, each data point
represents the system in thermal steady state. In the reference
sample which behaves as an ordinary SINIS cooler (i.e., with
negligible charging energy), there is a one-to-one correspondence
between the island electron temperature and the measured probe
voltage for a fixed current bias. The characteristics of the heat
transistor are drastically different, as seen in
Fig.~\ref{fig:data}(b).
In order to perform accurate thermometry on
the heat transistor sample with varying $V_g$ and $\VDS$, one needs
to determine the temperature $\TN$ by using a complete model
which takes into account the exact charge number distribution.

Steady state is reached at a temperature where the cooling power is
matched to the external heat load. The dominating source of heat
load into the island electrons is electron-phonon heat flux
($\Qdot_\mathrm{ep}$), which can be modeled by \cite{Roukes}
\begin{equation}
\Qdot_\mathrm{ep} = \Sigma \Vol (T_0^5 - \TN^5), \label{eq:Qeph}
\end{equation}
where $\Sigma$ is a material parameter, $\Vol$ is the volume of N
island, and $T_0$ is the phonon temperature. Using bulk values from
Ref.~\cite{Swartz_RMP} for the Kapitza resistance between the island
and substrate phonons, we estimate that the change in the
temperature of the island phonons due to electron cooling is less
than 10\% of the observed drop in the electron temperature. The true
Kapitza resistance for thin films is most likely even lower as one
can assume a common phonon system for the film and substrate
\cite{Dorozhkin}, and thus we neglect lattice cooling unlike in
Ref.~\cite{Sukumar}, and use $T_0 = \Tbath$. In our modeling, we
further ignore heating of the superconducting reservoirs due to hot
quasiparticles extracted from the island, whose effect is supposedly
weak due to relatively high junction resistances and moderate
cooling power. Consequently, we take $\TS = \Tbath$.
On the other hand, at sub-100 mK bath temperatures, electrons are
overheated due to noise via the electrical leads, and accurate
comparison with the theoretical model becomes difficult.

\begin{figure}
    \begin{center}
    \includegraphics[width=\columnwidth]{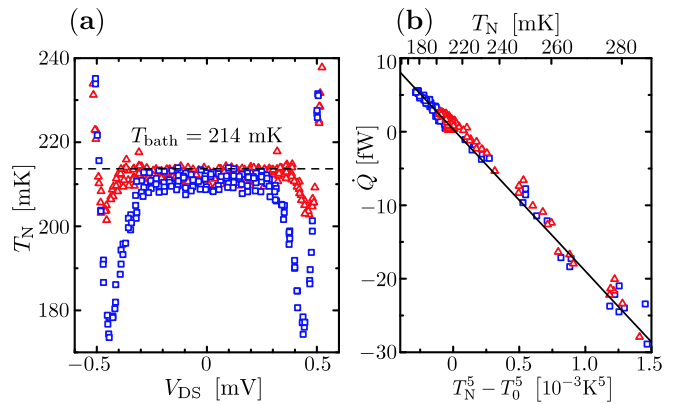}
    \end{center}
    \caption{(color online) (a) Electron temperatures extracted from the data acquired
    at bath temperature 214 mK with the gate open ($n_g$ = 1/2,
    blue squares) and closed ($n_g$ = 0, red triangles).
    (b) The theoretical cooling power vs. $\TN^5 - T_0^5$ for
    each data point of the left panel, and a linear fit to it. According to
    Eq.~(\ref{eq:Qeph}), the slope is given by $-\Sigma \Vol$.
    }
    \label{fig:temp}
\end{figure}

In analysis of the measurement data, we concentrate on
characterizing the cooling properties of the device at the extremal
gate positions.
These have to be identified as the local minima and maxima of either
$V_\txt{probe}$ or the drain-source current as the gate voltage is
swept.
In Fig.~\ref{fig:temp}(a), we present the electron temperatures
corresponding to open and closed gate positions derived from the
experimental data at bath temperature 214 mK. Similar results were
obtained at other bath temperatures as well, but only in the range
170 -- 250 mK we have both accurate thermometry and significant
temperature reduction. The temperature $\TN$ was evaluated
numerically for each data point using the experimentally determined
sample parameters, the observed values of $\Vprobe$ and $\VDS$, and
the set value of probe current, which was 3.2~pA in this
measurement. The electron temperatures obtained from the model for
small bias voltages $\VDS$, for which no cooling is expected, differ
less than 10~mK from the bath temperature for both open and closed
gate, demonstrating a good agreement between the experiment and
theory, and supporting the reliability of this method of
thermometry. With open gate, the transistor should function almost
like a regular SINIS cooler, and cooling peaks are indeed found at
bias voltages $\VDS \simeq \pm 2 \Delta/e$. Increasing the bias
voltage over $\sim 2\Delta/e$ brings the device rapidly into the
ohmic heating regime, which is observed as a sharp rise of the
electron temperature. For the closed gate configuration, the cooling
peaks are weaker and occur at a slightly larger bias voltage, as
tunneling electrons have to overcome both the Coulomb and the
superconducting gap.

A quantitative analysis of the observed temperature drops is
presented in Fig.~\ref{fig:temp}(b). The plot shows the numerically
calculated total cooling power versus $\TN^5 - T_0^5$, using the
$\TN$ values from Fig.~\ref{fig:temp}(a). According to the heat
balance equation $\Qdot = \Qdot_\mathrm{ep}$, the dependence should
be linear with the slope given by $-\Sigma \Vol$. Agreement with the
model is excellent throughout the whole data range from cooling
($\Qdot > 0$) to resistive heating ($\Qdot < 0$). Through a linear
fit we obtain $\Sigma = (2.3 \pm 0.3) \times 10^9 \unit{W K^{-5}
m^{-3}}$, the main source of uncertainty being the effective island
volume. The literature value $\Sigma = 2 \times 10^9 \unit{W K^{-5}
m^{-3}}$ \cite{Giazotto_RMP} is within the experimental error.
Estimates for $\Sigma$ obtained from data at other bath temperatures
are consistent with the above result, demonstrating the stability of
the method. Furthermore, the observed on/off ratio of cooling power
upon gate variation at the optimal bias point was as large as 3.2.

Figure~\ref{fig:data}(b) displays the $\Vprobe$ curves (blue lines)
calculated using the model which accounts for changes in the island
electron temperature with the experimentally obtained
 value $\Sigma = 2.3 \times 10^9 \unit{W K^{-5} m^{-3}}$.
For reference, we also show curves with the electron temperature
fixed to $T_\txt{bath}$ (red dashed lines), demonstrating which of
the observed features should be attributed to changes in
temperature. All the qualitative features of $\Vprobe$ are
reproduced, including the gentle bumps appearing at higher
temperatures, which are not indicative of cooling, but of changes in
the width of the charge number distribution.
These features can be observed in $\Vprobe$ when $\EC \lesssim \kB
\TN$, $\TN \approx \TS$ and when $\Vprobe$ is small. As expected,
significant differences between the cooled $\TN$ and $\TN \equiv
\Tbath$ curves appear only at low temperatures and with open gate,
whereas with closed gate, features due to change in the island
electron temperature remain always small. The measured amplitude of
gate modulation at $\Tbath = 137\unit{mK}$ is slightly smaller than
expected. The difference can be attributed to heat load from
electrical noise, to which the measurement is increasingly sensitive
at low temperatures. The smallness of the cooling power of the probe
current can be verified by the almost vanishing difference between
the theoretical curves for constant and variable $\TN$ near $\VDS =
0$. At high bath temperatures, the measured amplitude of the gate
modulation and the shape of the envelope curves are as expected,
although there are unexplained offsets of the order of $10\unit{\mu
V}$ in the measured data at different bath temperatures as compared
to theoretical curves. We have also verified that the theoretical
model reproduces accurately the measured cooler junction $I$-$V$
curves at all bath temperatures (data not shown), which further
validates the accuracy of the model and the fitted sample
parameters.

In conclusion, we have investigated experimentally and theoretically
the thermal DC properties of a Coulomb-blockaded SINIS structure. We
have demonstrated that the system operates as a mesoscopic heat
transistor, \ie, it shows a gate modulation of heat flux. In
particular, cooling power modulation as large as 70\% of its maximum
has been observed. The agreement between the experiment and a theory
based on Coulomb-blockaded single-electron tunneling is compelling.
Furthermore, we have also proved that an NIS probe can be used as a
sensitive thermometer even in the presence of charging effects,
provided that the charge number distribution on the normal metal is
carefully modeled. The influence of this distribution can be reduced
up to large extent by selecting the bias point properly. Finally,
our work might provide guidelines for the investigation of heat
transport at the mesoscopic scale, e.g., to either avoid or make use
of the effects arising from single-electron charging in electronic
refrigeration.

We acknowledge the NanoSciERA project ''NanoFridge'' of the EU  and
the Academy of Finland for financial support.

\bibliographystyle{apsrev}
\bibliography{thesis}

\end{document}